\documentstyle[epsfig,eqsecnum,aps]{revtex}
\begin{document}
\draft
\title{Nuclear shape dependence of Gamow-Teller distributions
in neutron-deficient Pb isotopes}

\author{P. Sarriguren, O. Moreno, R. \'Alvarez-Rodr\'{\i}guez, and 
E. Moya de Guerra}
\address{Instituto de Estructura de la Materia,
Consejo Superior de Investigaciones Cient\'{\i }ficas, \\
Serrano 123, E-28006 Madrid, Spain}

\date{\today}
\maketitle

\begin{abstract}

We study Gamow-Teller strength distributions in the neutron-deficient
even isotopes $^{184-194}$Pb in a search for signatures of deformation.
The microscopic formalism used is based on a deformed quasiparticle
random phase approximation (QRPA) approach, which involves a self-consistent 
quasiparticle deformed Skyrme Hartree-Fock (HF) basis and residual 
spin-isospin forces in both the particle-hole and particle-particle
channels. By analyzing the sensitivity of the Gamow-Teller strength
distributions to the various ingredients in the formalism, we conclude
that the $\beta$-decay of these isotopes could be a useful tool to look
for fingerprints of nuclear deformation.

\end{abstract}


\section{Introduction}

Neutron-deficient Pb isotopes turn out to be a unique laboratory to explore
the phenomenon of shape coexistence in nuclei and have recently been a subject 
of much experimental and theoretical interest. The existence of at least one
low-lying excited $0^+$ state in all even-even Pb isotopes between $A=184$
and $A=194$ has been experimentally observed at excitation energies below
1 MeV (see Ref. \cite{julin} and references therein). The most extreme cases
are $^{186}$Pb and $^{188}$Pb with two excited $0^+$ states below 700 keV
\cite{heese,andreyev}. These are good examples where the ground state and
the first two excited states, which are below 1 MeV, correspond to
macroscopical shapes. Experimental data on the excited states of these Pb
isotopes are not just limited to a few $0^+$ states. Rotational bands built
on top of these $0^+$ states have also been observed \cite{julin,dracoulis}, 
whose properties have been used to interpret them as corresponding to oblate 
and prolate deformations.

The existence of the $0^+$ low-lying states arises from the combined effect 
of a proton shell gap at $Z=82$ and a large number of hole neutrons in the 
$N=126$ core. The influence of the magic proton number $Z=82$ is particularly 
strong and the ground state of Pb isotopes is known to be spherical down to 
$^{182}$Pb. The weakening of the magicity of the $Z=82$ shell manifests 
itself through the appearance of low-lying $0^+$ states.

Different types of models have been invoked to explain the coexistence of
several $0^+$ states at low energies. In a shell model picture, the excited
$0^+$ states are interpreted as two-quasiparticle and four-quasiparticle
configurations \cite{heyde1}. Protons and neutrons outside the inert core
interact through pairing and quadrupole interactions to generate deformed
structures. Within a mean field approach, this is interpreted as a
macroscopic change from a spherical to a deformed shape. The energy of the
different shape configurations can be calculated using a nuclear potential,
where the energies of the single particle orbitals depend on the deformation.
Potential energy surface calculations have become more and more sophisticated
with time, resulting in accurate descriptions of the nuclear shapes and the
configurations involved. Calculations based on phenomenological mean fields
and Strutinsky method \cite{bengtsson,naza}, predict the existence of several
competing minima in the deformation energy surface of neutron-deficient Pb
isotopes. Self-consistent mean field calculations \cite{smirnova,niksic02}
and calculations including correlations beyond mean field 
\cite{duguet,egido,bender04} confirm these results.

Sophisticated configuration mixing calculations \cite{duguet,egido,bender04}
have been shown to support the mean field result that these neutron deficient
systems display more than one minimum corresponding to spherical and deformed 
shapes. The lowest excited states in configuration mixing calculations have
average deformations close to the deformation of the mean field minima 
\cite{duguet,egido,bender04}. All the approaches analyzed in Ref. \cite{egido} 
from mean field up to very sophisticated angular momentum projected generator 
coordinate method with Gogny forces provide the same underlying basic picture
of strong coexisting spherical, oblate and prolate shapes. Although the mean 
field approach does not reproduce quantitatively the experimental relative 
energies of the $0^+$ states, it provides a good qualitative description of 
the three minima and is a good approximation to associate the $0^+$ states 
observed at low energies with coexisting energy minima in the energy surface. 
The physical $0^+$ observed states have been also considered \cite{wood} as 
superpositions of spherical, oblate or prolate configurations, with relative 
weights determined phenomenologically, in the spirit of the shape mixing 
picture \cite{wood}. Such kind of calculations have been done in 
Ref. \cite{vanduppen} for $^{190-200}$Pb isotopes.

Mean field predictions of deformed ground states in the neutron-deficient 
Pb region are very sensitive to fine details in the calculations, specially to 
pairing effects. This has been shown for example within deformed relativistic 
mean field (RMF) calculations. Standard RMF calculations 
\cite{niksic02,yoshida94} do not reproduce the spherical ground
states in this mass region, being at variance with experimental data. In RMF, 
the ground states of neutron-deficient Pb isotopes are found to be deformed for
standard forces with constant pairing gaps \cite{mehta04}. However, it has
been shown \cite{yoshida97} that using constant strengths for the pairing
interaction (which makes the gap parameters strongly dependent on deformation),
produces spherical ground states for all Pb isotopes. More recently 
\cite{niksic02}, an improved pairing treatment has been used by means of
relativistic Hartree-Bogoliubov (RHB) calculations, using the NL3 effective
interaction and a finite range Gogny interaction to describe the pairing
properties. The results of these calculations show that $^{184}$Pb and 
$^{186}$Pb have spherical ground states with low-lying oblate and prolate
minima in qualitative agreement with experiment. However, with increasing
neutron number, the oblate minimum is lowered in energy and $^{188-194}$Pb
have oblate ground states, contradicting the experimental data \cite{julin}. 
In order to recover the spherical ground states, a new parametrization of
the effective interaction was proposed, which takes into account the sizes 
of spherical shell gaps.

The special sensitivity of the ground states in this mass region to 
deformation and pairing effects has been also studied from non-relativistic 
mean field calculations. In Ref. \cite{tajima1} it was found that the
potential energy curves in neutron-deficient Pb isotopes are very sensitive 
to the relative intensity of the neutron versus proton pairing strengths. 
In this reference it was shown that slightly different choices of the 
pairing strengths produce different ground state shapes.

On the other hand, it has been shown \cite{frisk,sarr} that the decay 
properties of unstable nuclei may depend on the nuclear shape of the
decaying parent nucleus. In particular, the Gamow-Teller (GT) strength
distributions corresponding to $\beta^+$-decay of proton-rich nuclei in the
mass region $A\approx 70$ has been studied systematically \cite{sarr} as a
function of the deformation, using a deformed HF+BCS+QRPA approach. 
In Ref. \cite{sarr} the GT strength distributions of several nuclei were 
identified to be particularly sensitive to the deformation of the 
$\beta^+$-emitter, and a comparison of the theoretical results with recent 
accurate measurements \cite{isolde} has been used to determine the nuclear 
shape in neutron deficient Kr and Sr isotopes.

In this paper we study the GT strength distributions of neutron-deficient
Pb isotopes from $A=184$ up to $A=194$ and their dependence on deformation.
The aim here is to identify possible signatures that allow one to determine
the existence of spherical, prolate or oblate nuclear shapes, in the spirit
of the shape coexistence picture, by looking at the $\beta^+$-decay pattern. 
Although possible effects of the quadrupole shape mixing cannot be discarded
and a QRPA-type approach might not be a good approximation to treat systems
undergoing large amplitude collective motions, the calculations performed in
this paper are perfectly reasonable as a first estimate of the $\beta^+$-decay
patterns. This study is timely because the possibility to carry out these 
measurements is being considered at present \cite{algora}.

The paper is organized as follows. In Sec. II we present a summary of the
theoretical framework used in our calculation. Sec. III contains our results
on the Gamow-Teller strength distributions of the Pb isotopes, as well as 
a discussion of their dependence on deformation. The conclusions are written
in Sec. IV.

\section{Theoretical framework}

In this Section we summarize briefly the theory involved in the microscopic
calculations. More details can be found in Ref. \cite{sarr}. The theoretical
formalism used here is based on a deformed Hartree-Fock mean field calculation,
using an effective two-body Skyrme interaction and including pairing
correlations in BCS approximation. The single particle energies, wave
functions, and occupation probabilities are generated from this mean field. 
We consider in this paper the force Sk3 \cite{sk3}, as an example of
one of the most simple, traditional and successful Skyrme interaction and the
force SG2 \cite{sg2}, which has been successfully tested against spin and 
isospin excitations in spherical \cite{sg2} and deformed nuclei 
\cite{sarr,noj}. Some results will also be given for the more recent force
SLy4 \cite{sly4}.

The formalism is safely restricted to axially deformed nuclear shapes because 
it has been shown \cite{bengtsson,bender04} that no features of particular 
interest, such as minima or flat regions, appear for triaxial shapes in 
neutron-deficient Pb isotopes. Time reversal and axial symmetry are then 
assumed here. The single-particle wave functions are expanded in terms of
the eigenstates of an axially symmetric harmonic oscillator in cylindrical
coordinates, using eleven major shells. We calculate the energy surfaces as
a function of deformation for all the even-even neutron-deficient Pb isotopes
under study here. For that purpose, we perform constrained HF calculations
with a quadrupole constraint \cite{constraint} and minimize the HF energy
under the constraint of keeping fixed the nuclear deformation. In order to
study the above mentioned sensitivity of the energy surfaces to pairing
correlations, we compare the results obtained using fixed gap parameters
for protons and neutrons with the results obtained using constant pairing
strengths $G_{\pi,\nu}$. The fixed gap parameters are determined
phenomenologically from the odd-even mass differences \cite{audi} when this
experimental information is available, otherwise, we use the expression
$\Delta = 12 A^{-1/2}$ MeV. The  $G_{\pi,\nu}$ strengths are determined
from the expression 

\begin{equation}
G_{\pi(\nu)}=\frac{18}{11+Z(N)} \, {\rm MeV}.
\label{gpn}
\end{equation}

To describe Gamow-Teller excitations we use the QRPA adding to the mean field 
a separable spin-isospin residual interaction, which is expected to be the most 
relevant interaction for that purpose. This interaction contains two parts, 
particle-hole ($ph$) and particle-particle ($pp$). The $ph$ part is responsible 
for the position and structure of the GT resonance \cite{sarr,moller,hir,homma} 
and it is a common practice to fit the strength of the force to reproduce the 
energy of the resonance \cite{homma}. The advantage of using
separable forces is that the QRPA energy eigenvalue problem is reduced to find 
the roots of an algebraic equation. The particle-particle interaction is a 
neutron-proton pairing force in the $J^\pi=1^+$ coupling channel. We introduce 
this interaction in the usual way \cite{hir,muto,engel}, that is, in terms of 
a separable force with a coupling constant $\kappa ^{pp}_{GT}$, which is 
fitted to the phenomenology. Since the peak of the GT resonance is almost 
insensitive to the $pp$ force, $\kappa ^{pp}_{GT}$ is usually adjusted to 
reproduce the half-lives \cite{hir,homma}.

The pnQRPA phonon operator for GT excitations in even-even nuclei is 
written as

\begin{equation}
\Gamma _{\omega _{K}}^{+}=\sum_{\pi\nu}\left[ X_{\pi\nu}^{\omega _{K}}
\alpha _{\nu}^{+}\alpha _{\bar{\pi}}^{+}-Y_{\pi\nu}^{\omega _{K}}
\alpha _{\bar{\nu}}\alpha_{\pi}\right]\, ,  
\label{phon}
\end{equation}
where $\pi$ and $\nu$ stand for proton and neutron, respectively,
$\alpha ^{+}\left( \alpha \right) $ are quasiparticle creation
(annihilation) operators, $\omega _{K}$ are the RPA excitation energies, 
and $X_{\pi\nu}^{\omega _{K}},Y_{\pi\nu}^{\omega _{K}}$ the forward and 
backward amplitudes, respectively. It satisfies
\begin{equation}
\Gamma _{\omega _{K}} \left| 0\right\rangle=0\, ; \qquad
\Gamma ^+ _{\omega _{K}} \left| 0\right\rangle = \left|
\omega _K \right\rangle .
\end{equation}
Solving the pnQRPA equations \cite{moller,muto},
the GT transition amplitudes in the intrinsic frame of even-even nuclei
connecting the QRPA ground state of the parent nucleus
$\left| 0\right\rangle $ to one phonon states in the daughter nucleus 
$\left| \omega _K \right\rangle $, are given by
\begin{equation} 
\left\langle \omega _K | \beta _K^{\pm} | 0 \right\rangle = \mp 
M^{\omega _K}_\pm \, ,
\label{amplgt}
\end{equation}
where  $\beta _K^{\pm}=\sigma_K t^{\pm},\quad K=0,\pm 1$, and 

\begin{equation}
M_{-}^{\omega _{K}}=\sum_{\pi\nu}\left( q_{\pi\nu}X_{\pi\nu}^{\omega _{K}}+
\tilde{q}_{\pi\nu}Y_{\pi\nu}^{\omega _{K}}\right) \, ; \qquad 
M_{+}^{\omega _{K}}=\sum_{\pi\nu}\left( \tilde{q}_{\pi\nu}
X_{\pi\nu}^{\omega _{K}}+ q_{\pi\nu}Y_{\pi\nu}^{\omega _{K}}\right) \, ,
\end{equation}
with
\begin{equation}
\tilde{q}_{\pi\nu}=u_{\nu}v_{\pi}\Sigma _{K}^{\nu\pi };\ \ \ 
q_{\pi\nu}=v_{\nu}u_{\pi}\Sigma _{K}^{\nu\pi}\, ;\qquad
\Sigma _{K}^{\nu\pi}=\left\langle \nu\left| \sigma _{K}\right| 
\pi\right\rangle \, ,
\label{qs}
\end{equation}
where $v'$s are occupation amplitudes ($u^2=1-v^2$).

Once the intrinsic amplitudes in Eq. (\ref{amplgt}) are calculated, the
Gamow-Teller strength $B(GT)$ in the laboratory frame for a transition 
$I_i K_i (0^+0)\rightarrow I_fK_f(1^+K)$
can be obtained as

\begin{equation}
B^{\pm}(GT)= \sum_{M_i,M_f,\mu} \left| \left< I_fM_f \left| \beta ^\pm _\mu
\right| I_i M_i \right> \right|^2= \left\{ \delta_{K_f,0} \left< \phi_{K_f} 
\left|  \beta ^\pm _0 \right| \phi_0\right> ^2 +2\delta_{K_f,1} \left< 
\phi_{K_f} \left|  \beta ^\pm _1 \right| \phi_0\right> ^2 \right\} \, ,
\label{streven}
\end{equation}
in units of $g_A^2/4\pi$. To obtain this expression we have used the initial
and final states in the laboratory frame expressed in terms of the intrinsic
states $|\phi_K >$, using the Bohr-Mottelson factorization \cite{bm}.

The $\beta$-decay half-life is obtained by summing up all the allowed 
transition probabilities, weighed with phase space factors, up to 
states in the daughter nucleus with excitation energies lying within 
the $Q$-window,   

\begin{equation}
T_{1/2}^{-1}=\frac{A^2}{D}\sum_{\omega }f\left( Z,\omega \right) B(GT)\, ,
 \label{t12}
\end{equation}
where $f\left( Z,\omega \right) $ is the Fermi integral \cite{gove} and 
$D=6200$~s. We include standard effective factors

\begin{equation}
A^{2}=\left[ \left( g_{A}/g_{V}\right) _{\rm eff}\right] ^{2}=\left[
0.7\left( g_{A}/g_{V}\right) _{\rm free}\right] ^{2} \, . \label{quen}
\end{equation}
In $\beta^+/EC$ decay, $f\left( Z,\omega \right) $ consists of two parts,
positron emission and electron capture. In this work we have computed them 
numerically for each value of the energy, as explained in Ref. \cite{gove}.

\section{Results and discussion}

In this section we present the results obtained from the above mentioned
theoretical formalism for the $\beta$-decay patterns of the neutron-deficient
$^{184,186,188,190,192,194}$Pb isotopes. First, we discuss the energy surfaces
and shape coexistence expected in these isotopes. Then, we show on some 
examples the sensitivity of the Gamow-Teller strength distributions to the 
various ingredients in the calculation. Finally, we present the results 
obtained for the Gamow-Teller strength distributions with special attention
to their dependence on the nuclear shape and discuss their relevance as 
signatures of deformation to be explored experimentally.

\subsection{Deformation energy curves}

We show in Fig. 1 the HF energies calculated with the Skyrme force Sk3 and
pairing correlations treated in the fixed gap approach, for all the isotopes
considered, as a function of the quadrupole deformation parameter $\beta$.
The latter is defined as 

\begin{equation}
\beta = \sqrt{\frac{\pi}{5}}\frac{Q_p}{Z<r^2>}
\end{equation}
in terms of the proton quadrupole moment $Q_p$ and the charge r.m.s. radius
$<r^2>$. For a better comparison, the curves are scaled to the energy of 
their ground states and are shifted by 1,2,3,4,5 MeV for A=186,188,190,192,194,
respectively. As we can see in the figure, prolate solutions corresponding to
values ($\beta > 0$) are always present from A=184 to A=190, and are indeed
the ground state for this choice of force and gap parameters. The prolate
minimum shows only a shoulder in A=192 and disappears completely in A=194,
where only the spherical solution remains. Spherical local minima can be
seen in all the isotopes except for A=192, the one in A=190 being very shallow.
Oblate minima ($\beta < 0$) are also present in all the isotopes except for the
one closer to stability A=194. We can also see in Fig. 1 that the oblate
deformation corresponding to the local oblate minima in all the isotopes is
rather stable at $\beta = -0.2$, while the prolate deformation at the minima
becomes larger as we remove more and more neutrons, to reach $\beta=0.32$ in
A=184. Mass deformations follow closely charge deformations.

Our results are in qualitative agreement with experiment in the sense that
we find different coexisting shapes in this mass region, although they are
at variance with experiment \cite{julin} in the predicted ground state shape. 
Experimentally, the ground states of all the isotopes considered here are 
found to be spherical, while we find a spherical shape as the ground state in 
A=194, an oblate shape in A=192 and a prolate shape in the remaining cases 
considered (A=184-190). This discrepancy is not surprising because as we have 
mentioned in the Introduction, even the more sophisticated formalisms fail 
in some instances \cite{niksic02} to account for the spherical ground state, 
showing an extreme sensitivity to fine details of the two-body interactions.
To illustrate this sensitivity, we compare in Fig. 2 on the example of A=184,
the energy deformation curves obtained when we change the force from Sk3 to
SG2 and SLy4, and when we change the pairing correlations from fixed gap 
parameters to constant pairing interaction with $G_{\pi(\nu)}$ strengths 
given by Eq. (\ref{gpn}). We can see in Fig. 2 that although we get a prolate 
ground state in all the cases considered, the calculation with a constant 
pairing strength (dotted line) takes the spherical minimum very close in 
energy to the ground state. Indeed by slightly changing the $G$-values (the 
relative values of $G_{\pi}$ versus $G_{\nu}$), we can get the spherical 
solution as the ground state. An analysis along these lines was made in 
Ref. \cite{tajima1} with analogous results. It is also worth mentioning 
that the force SLy4 produces three minima which are very close in energy.

The results shown in Fig. 2 represent what happens when using other forces 
and other gap parameters or pairing strength values. That is,
the deformation minima remain at about the same deformation values but their
relative energies may change considerably. This example is also representative
of the situation with respect to other isotopes.

\subsection{Gamow-Teller strength distributions and deformation}

In this subsection we study the energy distribution of the Gamow-Teller 
strengths calculated at the equilibrium shapes that minimize the energy
of the nucleus. First, we settle the values of the coupling strength
constants of our separable residual $ph$ and $pp$ forces. In this work 
we use the same value for all the isotopes considered, which is fixed by
comparison to the experimental $B^-(GT)$ strength resonance observed from
(p,n) charge exchange reactions in $^{208}$Pb at an energy of 19.2 MeV 
relative to the parent nucleus $^{208}$Pb \cite{gaarde}. In this way,
we reproduce the energy of the resonance with 
$\chi ^{ph}_{GT}= 0.09\, (0.07)$ MeV, when using SG2 (Sk3) forces and these 
are the values we use here. Once the value of the $ph$ spin-isospin residual
force has been established, the value of the $pp$ residual force is usually
chosen to reproduce the measured half-life. In the present case, we have
calculated the half-lives as a function of the $\kappa ^{pp}_{GT}$ parameter
and have found that they are very little dependent on this parameter. This
is illustrated in Fig. 3, where we can see the results obtained for $T_{1/2}$
from Eqs. (\ref{t12}) and (\ref{quen}). The results correspond to the case
$^{190}$Pb with an experimental half-life $T_{1/2}=71$ s. The effect of the
$pp$ force on the GT strength distributions can be seen in the inset, where
we have plotted the distributions for the oblate and prolate shapes with  
$\kappa ^{pp}_{GT}=$ 0 and 0.02 MeV. From now on, we use 
$\kappa ^{pp}_{GT}=0.02$ MeV in the calculations of the GT strengths.
This value is compatible with the parametrization of Ref. \cite{homma},
where it was found $\kappa ^{pp}_{GT}=0.58/A^{0.7}$ MeV.

In Fig. 4 we study the influence of both the Skyrme force and the pairing
treatment used on the GT strength distributions. We do that on the example
of the isotope $^{184}$Pb and perform the calculations for the different
shapes that minimize the energy of this isotope, which are shown in
Fig. 2. The purpose of this figure is to show that the different profiles
obtained for different deformations remain when the force and/or pairing 
are changed. This implies that the GT profile is characteristic of the shape
and depends little on the details of the two-body forces used. 
Fig. 4, as well as all the coming figures containing GT strength distributions,  
shows in addition to the discrete
strengths, a continuous distribution resulting from a folding procedure
using $\Gamma=1$ MeV width gaussians on the discrete spectrum. The $B(GT)$
strength is plotted versus the excitation energy of the corresponding
daughter nucleus. On the left panels in Fig. 4 we show the results from
the oblate shape, in the middle the results from the spherical shape, and
on the right panels we show those from the prolate shape. For an easier
comparison between the various results, we have used in all the panels the
same scales of strengths and energies. In the upper row we compare the
results from two different Skyrme forces Sk3 and SG2, using pairing
correlations with fixed gap parameters $\Delta_\pi = \Delta_\nu=0.9$ MeV,
as obtained from the general expression given in Sec. 2. Results from Sk3 
and SG2 are plotted upward and downward, respectively. 
In the lower row we compare the results from Sk3 using fixed gap
parameters (plotted upward) and constant pairing strengths given by 
Eq.(\ref{gpn}) plotted downward. The main point to stress in this figure
is the fact that the structure of the GT strength profiles is different
for each deformation but very similar when we compare the calculation of
reference (Sk3 with fixed pairing gaps) plotted upward with those plotted
downward. This is specially true in the low energy region where the oblate
shape gives rise to a three bump structure below 5 MeV, the spherical
shape a single bump structure, and the prolate shape a two peaked structure
with a first strong single peak at very low energy and a more spread second
bump at higher energies.

The main difference between the calculations from Sk3 and SG2 is a small
shift to lower excitation energies in the case of SG2. This is better seen
in the spherical case where a single peak is present. Actually, in the strict
spherical limit the fragmentation of the strength observed in the figure
would collapse to single excitation energies corresponding to the transitions
between degenerate spherical orbitals. The fragmentation observed arises from
the small deformation present in our deformed formalism ($\beta = 0.003$),
which is enough to break the degeneracy of the third components of angular
momentum. The origin of the shift in the excitation energy observed with
different forces, can be traced back to the predicted energies for the
$h_{9/2}$ and $h_{11/2}$ spherical shells. The main contribution to the GT
strength at low energies comes from the transition connecting the almost
fully occupied $\pi h_{11/2}$ shell with the partially unoccupied
$\nu h_{9/2}$ shell, and the relative position of the two shells (closer
for SG2 than for Sk3) determines the GT excitation energy. This effect
translates into the deformed cases where the non-degenerate orbitals
spread around the energy of the spherical shell. The difference between
the calculations with different pairing treatments shown in the lower
panels is due to the different occupation probabilities of the orbitals
that weigh differently the matrix elements connecting the proton and
neutron states, as well as to the different quasiparticle energies that
finally shift the excitation energy of the GT transitions.

In Fig. 5 we show the results for the $B(GT)$ strength distributions in
($g_A^2/4\pi$) units, calculated within QRPA based on the force Sk3 with
fixed pairing gaps and residual separable interactions $ph$ and $pp$ 
with coupling strengths given by $\chi ^{ph}_{GT}=0.07$ MeV and 
$\kappa ^{pp}_{GT}=0.02$ MeV, respectively. In the left panels we show the
results obtained with spherical shapes from $^{184}$Pb up to $^{192}$Pb,
the middle panels are for oblate shapes and the right panels are  for prolate
shapes in the same chain of isotopes. In each column we have used the same
scale for the GT strength to appreciate better the changes as we increase the
number of neutrons. The scale on energies is always 7 MeV, enough to include
all the $Q_{EC}$ energies, whose maximum value is $Q_{EC}=5.84$ MeV in
$^{184}$Pb. We can see in this figure the progress of the profiles as we
change the mass number (number of neutrons). We can see a continuous
transition from the lightest (A=184) to the heaviest (A=192) isotopes,
that can be guided by the dashed lines crossing the figures.
Thus, we can see that the strength in the spherical case is accumulated
in a resonance peaked at around 3 MeV and quite stable against the number of
neutrons. This strength has its origin in the 
$\pi h_{11/2} \rightarrow \nu h_{9/2}$ transition. In the case of oblate
shapes, we observe one peak at low energies around 2 MeV, which stays in all
the isotopes although is weaker as we approach the stability. There is also
a second peak clearly seen in $^{184}$Pb that moves to lower energy with
mass and merges with the first peak gradually. Finally, we observe a third
peak that again moves to lower energy with mass. While this third peak is
practically the same in all the isotopes, the two first peaks become weaker
with increasing neutron number. An analysis of the structure of these
transitions reveals that while the first two peaks arise from several
transitions between negative parity states (mainly from 
$\pi h_{11/2} \rightarrow \nu h_{9/2}$ transitions), the third peak at higher 
energy corresponds to transitions between positive parity states with a
dominant transition $\pi (13/2)^+ \rightarrow \nu (11/2)^+$, connecting the 
proton $i_{13/2}$ shell with the neutron $i_{11/2}$ shell. 
The prolate cases show also one
first peak very strong and stable at around 1 MeV, originated by a single
transition, as we can see in the next figures. There is also a second peak
at higher energy in $^{184}$Pb. This peak is more fragmented than the lower
energy peak and moves to lower energy loosing strength as we move downward
in the figure. In the prolate case both peaks are originated from transitions
between negative parity states, reminiscent of the transition 
$\pi h_{11/2} \rightarrow \nu h_{9/2}$ in the spherical case.

Now, we can particularize the discussion case by case analyzing in detail each
isotope with regard to its deformation signatures. In the next figures 
(Figs. 6-11) we show for each isotope a comparison of the GT distributions
obtained from the spherical (upper panel), oblate (middle panel) and prolate
shapes (lower panel). We also show the discrete spectra, not folded, as they
come directly from the QRPA calculations.

In Fig. 6 for $^{184}$Pb we can see a bump between 2.5 and 3 MeV in the
spherical case. In the oblate case we get a three bump structure with peaks 
centered at 1.5, 3, and 5 MeV, respectively. The two peaks at lower energy
are fragmented within a range of 1 MeV while the third peak is little
fragmented. On the contrary, in the prolate case we see a very strong single 
peak at 1 MeV
and a second narrow peak at 3 MeV followed by a tail spread up to 5 MeV.
Similarly, we can see in Figs. 7-10 the following characteristics which are
common to all the isotopes: i) an accumulation of the strength between 
2.5 MeV and 3 MeV, depending on the isotope, for the spherical shape; 
ii) a very fragmented bump structure spread from 1.5 MeV up to 3.5 MeV, 
as well as a single peak located between 4 MeV and 5 MeV, depending on the 
isotope, in the oblate case; 
iii)  a single very strong peak at very low energy and a more spread structure
beyond 3 MeV in the prolate case. Fig. 11 shows the results for $^{194}$Pb, 
where only a stable solution, which is spherical, is found with Sk3 (see 
Fig. 1). The strength in this case is located beyond the $Q_{EC}$ value and 
therefore of no significance from the $\beta$ decay point of view. 
Our calculation with Sk3 predicts a stable isotope although the half-life 
has been measured to be $T_{1/2}=720$ s. We shall come back to this point 
later on when discussing our results for half-lives.

Now we focus our attention on the isotopes $^{188,190,192}$Pb, because
several features combine to make them the best candidates for $\beta$-decay
strength distributions measurements \cite{algora}. These features are:
i) a dominant $\beta^+/EC$ decay component (see Table 1);
ii) a relatively large half-life ($T_{1/2}=25-210$ s); and iii) a large enough
$Q_{EC}$ energy (3.3-4.5 MeV). In the case of
$^{188}$Pb, practically all the strength is found below its $Q_{EC}$ value
except for one peak at 4.5 MeV in the oblate case. A signature of deformation
in this isotope would be the existence of GT strength below an excitation
energy of 2 MeV, a region where the spherical shape does not generate any
strength. If a very strong narrow peak is found in this low energy region,
this would be an indication of a prolate shape, while if very fragmented
strength is found this would indicate an oblate shape. The situation is
similar in  $^{190}$Pb, but here it is even more evident the
existence of three regions. Appearance of strength below 2.5 MeV would
indicate again a deformed nuclear shape. If the strength were located below
1.5 MeV in the form of a narrow peak, it would indicate a prolate shape while 
its absence would rule out this possibility. In the case of $^{192}$Pb, part of
the strength is already located beyond the $Q_{EC}$ window but still we get
very clean signatures of deformation. The existence of a narrow strength
distribution very close to the ground state would clearly indicate a
prolate shape of the decaying nucleus and the absence of this strong peak
together with the existence of strength extended up to 3 MeV indicates an
oblate shape. The lack of strength below 2.5 MeV would rule out a deformed
shape.

We have also calculated the half-lives (Eq. (\ref{t12})) corresponding to
the $\beta^+ /EC$ decays of these isotopes with the purpose of contrasting
our results against the measured half-lives and checking that there is no 
fundamental disagreement with experiment. The $\beta^+ /EC$ decay mode 
competes in this mass region with $\alpha$-decay and the relative 
contribution of the two decays can be seen in Table 1 as the percentage of 
the $\beta^+ /EC$ decay with respect to the total decay \cite{audi}.
We can also see in Table 1 the experimental half-lives \cite{audi}, as
well as the calculated half-lives from the spherical, oblate and prolate
shapes with the forces Sk3 and SG2, using quenching factors given in
Eq. (\ref{quen}) and experimental $Q_{EC}$ values.
In principle, different $Q$-values should be used in the calculations
for different starting/ending energy minima. Nevertheless, the $Q$-values
calculated from the binding energies of parent and daughter are quite
similar for the various shapes and close to the corresponding experimental
values. We have also checked that the half-lives calculated with the
experimental $Q_{EC}$ values change in less than a factor of two
when $Q_{EC}$ is varied within 1 MeV around this value.  
This is at least the case in the more relevant isotopes $^{190,192}$Pb,
where the $\beta^+$ percentage is above $99 \%$. One should also mention
that the spherical cases are more sensitive to the $Q_{EC}$ values 
because the location of the GT resonance close to the $Q_{EC}$ value
may be critical. Nevertheless, the half-lives in the spherical cases are 
always much larger than the corresponding to the deformed cases.

The decay of the most unstable isotopes $^{184,186}$Pb are dominated by
$\alpha$-decay and therefore the predictions for the $\beta^+ /EC$ 
decay half-lives should be above the measured value. This is indeed 
observed in the table. In the case of $^{184}$Pb, the $\beta^+$ 
percentage is small ($20(15) \%$). Then, we expect the calculated 
$\beta^+ /EC$ half-lives to be roughly about one order of magnitude larger
than the total experimental half-life and this is observed in the table.
For $^{186}$Pb the decay is shared by $\alpha$ and $\beta $ 
emission in comparable proportions. Thus, we expect $\beta^+ /EC$ half-lives 
of the order of twice the total half-life measured, and this is roughly the 
case in Table 1. We also
observe a general characteristic common to all the isotopes, the
spherical shape produces always the largest half-life and the prolate
shape the lowest one with the oblate half-life in between. The most
interesting cases are $^{188,190,192}$Pb. These decays are clearly
dominated by $\beta^+ /EC$ decay with half-lives ranging from 
$T_{1/2}= 25.5$ s up to $T_{1/2}=210$ s. In $^{188}$Pb the experimental
$T_{1/2}=25.5$ s, is compatible with the prolate and oblate shapes in
both Sk3 and SG2. In $^{190}$Pb the experimental $T_{1/2}=71$ s, is much
closer to the oblate calculation ($T_{1/2}= 83$ s) and about a factor of
two larger than the prolate half-life ($T_{1/2}=27$ s) in Sk3. The SG2
results are in this case below experiment. The measured half-life of
$^{192}$Pb ($T_{1/2}=210$ s) is well reproduced by the oblate shape, but
it is a factor of 8 larger than the half-life of the prolate shape in Sk3. 
The results from SG2 are a factor of 2 (3) below experiment for the oblate 
(prolate) shape. In the case of $^{194}$Pb, as already indicated, the Sk3 
force predicts only one minimum, which is spherical and stable against 
$\beta$-decay. The oblate shape corresponding to the force SG2 predicts 
a half-life that agrees nicely with experiment, while the half-life
obtained with the spherical shape is too large.

As a general comment we can mention that the $\beta^+ /EC$ half-lives 
predicted by the spherical shapes are in general much larger than the
experimental half-lives.
As a consequence, even though the spherical shape corresponds experimentally
to the ground state of the nucleus, according to our calculations, the 
$\beta^+ /EC$ half-lives are determined to a large extent by the 
$\beta^+ /EC$ decay of the oblate and prolate shape isomers. Within the
present calculations, we would be unable to reproduce the half-lives if
we would restrict ourselves to spherical shapes.

\section{Conclusions}

We have studied the energy distribution of the Gamow-Teller strength in 
the neutron-deficient $^{184,186,188,190,192,194}$Pb isotopes. The study 
has been done within a deformed pnQRPA formalism with spin-isospin
$ph$ and $pp$ separable residual interactions. The quasiparticle mean field
includes pairing correlations in BCS approximation and it is generated from
a deformed HF approach with two-body Skyrme effective interactions. 

The parameters of the Skyrme forces are previously
determined by fits to nuclear matter and ground state properties of spherical
nuclei and the pairing gap parameters are extracted from the experimental masses.
The equilibrium deformation is derived self-consistently within the HF procedure. 
The coupling strength of the $ph$ residual interaction is fixed to reproduce
the energy of the GT resonance in $^{208}$Pb. The value obtained is close to 
the coupling strength derived consistently from the Skyrme force. Finally,
the coupling strength of the $pp$ residual interaction is taken from existing
global parametrizations, although we find our results to be quite insensitive 
to this value. Our calculation has basically no free parameters to 
fit locally any of the particular isotopes discussed.

Although the present procedure to generate the structure properties may 
be enough for a first estimate of the GT strength distributions, it would 
be interesting in the future to improve this description and discuss the 
effect of using the most recent Skyrme forces available in the literature 
together with a treatment of the pairing correlations using zero-range 
density dependent pairing interactions, where issues such as renormalization 
or regularization of the pairing force can be analyzed in this context. 

We have investigated the sensitivity of our results to the forces used 
and we arrive to the following conclusions:
i) In a given isotope, the energy deformation curve and, in particular, the 
relative energies of the spherical, oblate, and prolate minima, 
are very sensitive to the Skyrme and pairing forces used. 
The ground state shape predicted may be modified with different choices for 
these forces. Nevertheless, the qualitative behavior of the energy profile 
in each isotope remains unchanged against changes in the Skyrme and pairing 
forces in the
sense that the deformations at which the minima occur are hardly shifted and 
that the shape coexistence between spherical, oblate and prolate shapes is a 
constant feature in the more neutron deficient isotopes.
ii) The GT strength distributions calculated at the equilibrium deformations show
specific characteristics for each deformation that remain against changes of
the Skyrme and pairing forces. The effect of the deformation on the GT strength
distributions is much stronger than the effects coming from the
Skyrme or pairing forces used.
iii) As a consequence, we have been able to identify clear signatures of
deformation on the GT strength distributions of the Pb isotopes studied.
These signatures are related to the profiles of the GT strength distributions, 
which are peaked at different energies depending on the shape of the decaying
nucleus. A detailed analysis of the best candidates to be measured, 
$^{188,190,192}$Pb, has been performed and the characteristic features of 
their GT strength distributions have been identified.

The calculated $\beta^+ /EC$ half-lives show a strong dependence on the shape
of the nucleus. The spherical shape produces always the largest half-life, which
is far away from experiment. The prolate shape gives the lowest one while the 
oblate half-life lies in between and agrees better with experiment. The 
conclusion from our calculated half-lives is that in order to reproduce the 
experimental $\beta^+ /EC$ half-lives, the decay of the spherical isomer is 
not enough and one has to consider the decay of the oblate and prolate shape 
isomers as well.

\begin{center}
{\Large \bf Acknowledgments} 
\end{center}
We are grateful to A. Algora and B. Rubio for useful discussions.
This work was supported by Ministerio de Educaci\'on y Ciencia (Spain) under 
contract number BFM2002-03562. Two of us (O.M. and R.A.-R.) thank Ministerio
de Educaci\'on y Ciencia (Spain) for financial support.

\newpage


\begin{center}

\begin{table}[t]
\caption{ Percentage of the $\beta^+/EC$ involved in the total decay of the
Pb isotopes, experimental \protect\cite{audi} and theoretical half-lives [s],
obtained with the forces Sk3 and SG2. Experimental $Q_{EC}$ values have been
used.}
\label{table.1}
\begin{tabular}{rcccccccc}\cr
 A  & \% ($\beta^+/EC$) & $T_{1/2}^{\rm total}$ exp &  \multicolumn{3}{c} 
{$T_{1/2}^{\beta^+}$ (Sk3)}  & \multicolumn{3}{c} {$T_{1/2}^{\beta^+}$ (SG2)}  \cr
\cline{4-6} \cline {7-9} \cr
&&& oblate & prolate & spherical & oblate & prolate & spherical \cr
\hline
\cr
194 & 100        & 720 (30)   &  -   &   - & stable & 642 &   - & 2053   \cr
192 & 99.9941 (7)& 210 (6)    & 207  & 28  &   4385 & 132 &  84 & 763 \cr
190 & 99.60 (4)  & 71 (1)     & 83   & 27  &    344 &  53 &  12 & 195 \cr
188 & 90.7 (8)   & 25.5 (1)   & 32   & 18  &     85 &  24 &  22 & 62  \cr
186 & 60 (8)     & 4.82 (3)   & 15.4 & 7.7 &     47 & 8.2 & 6.2 & 16  \cr
184 & 20 (15)    & 0.490 (25) & 6.4  & 3.3 &     22 & 4.5 & 3.2 & 6.7
\cr
\end{tabular}
\end{table}

\newpage

\begin{figure}[t]
\epsfig{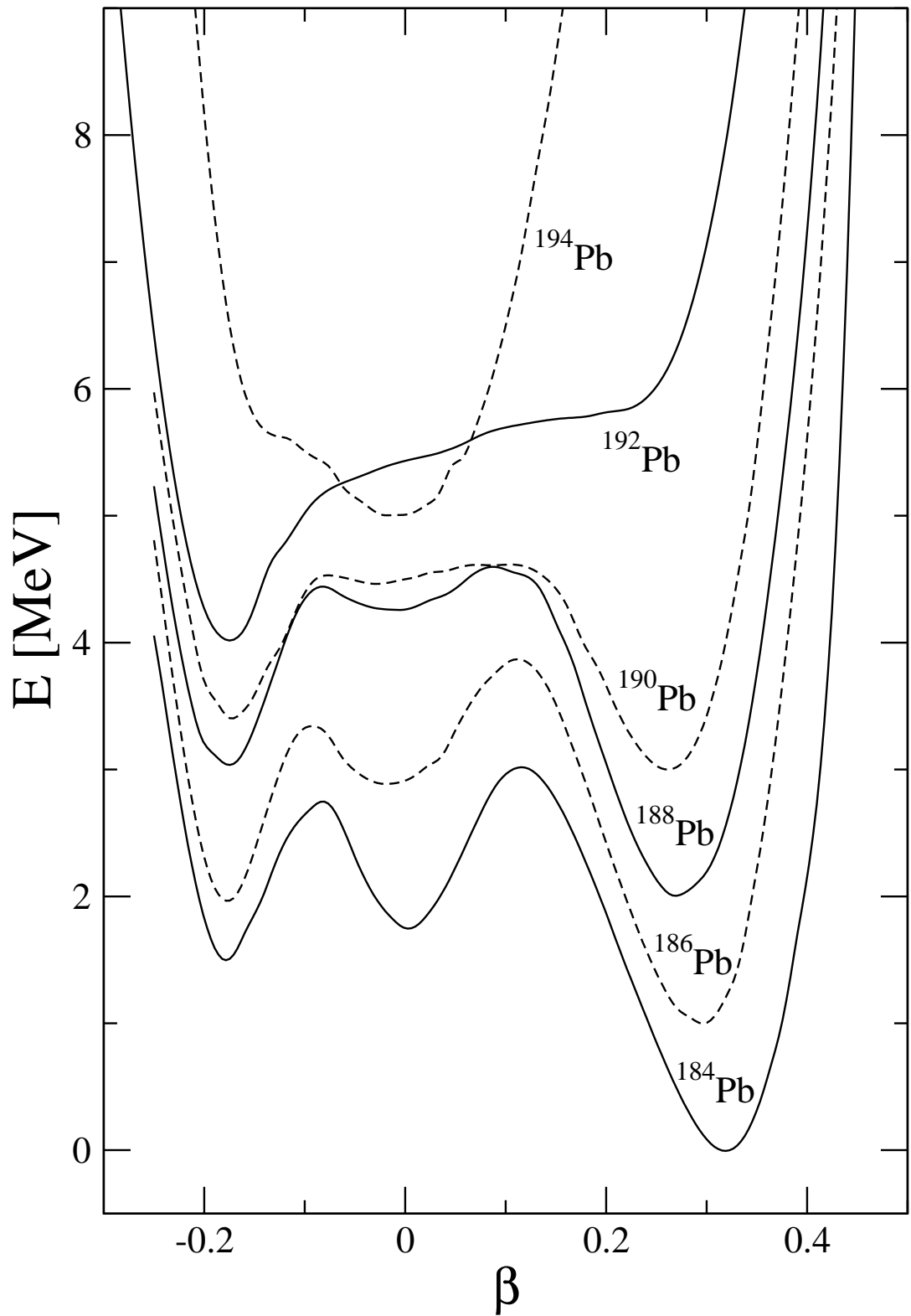}
\vskip 1cm
\caption{HF energy of the isotopes $^{184,186,188,190,192,194}$Pb obtained 
from constrained HF+BCS calculations with the force Sk3 and fixed pairing gap
parameters as a function of the quadrupole deformation $\beta$.}
\end{figure}

\newpage

\begin{figure}[t]
\epsfig{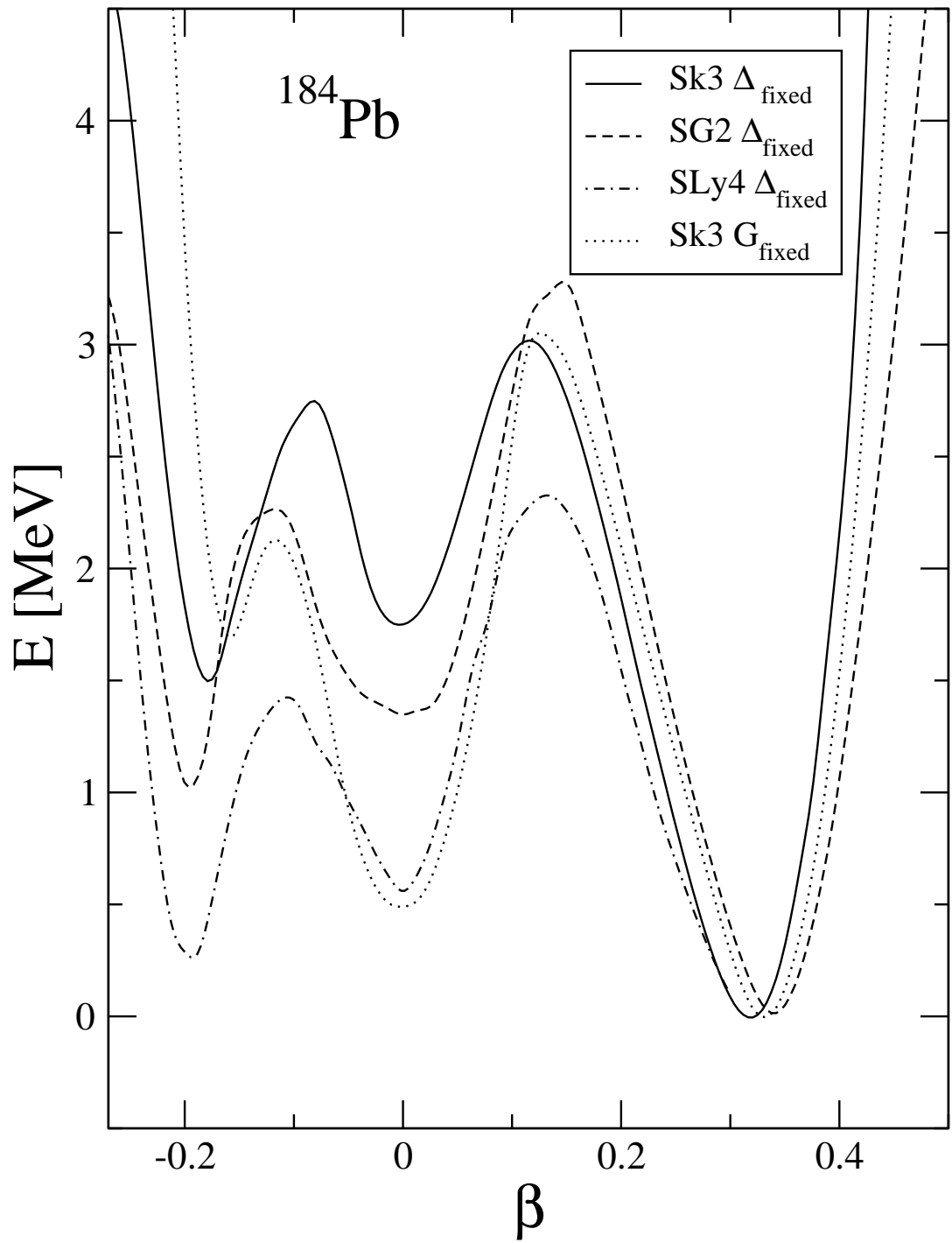}
\vskip 1cm
\caption{HF energy of $^{184}$Pb as a function of the quadrupole deformation 
$\beta$, for various forces and pairing treatments.}
\end{figure}

\newpage

\begin{figure}[t]
\epsfig{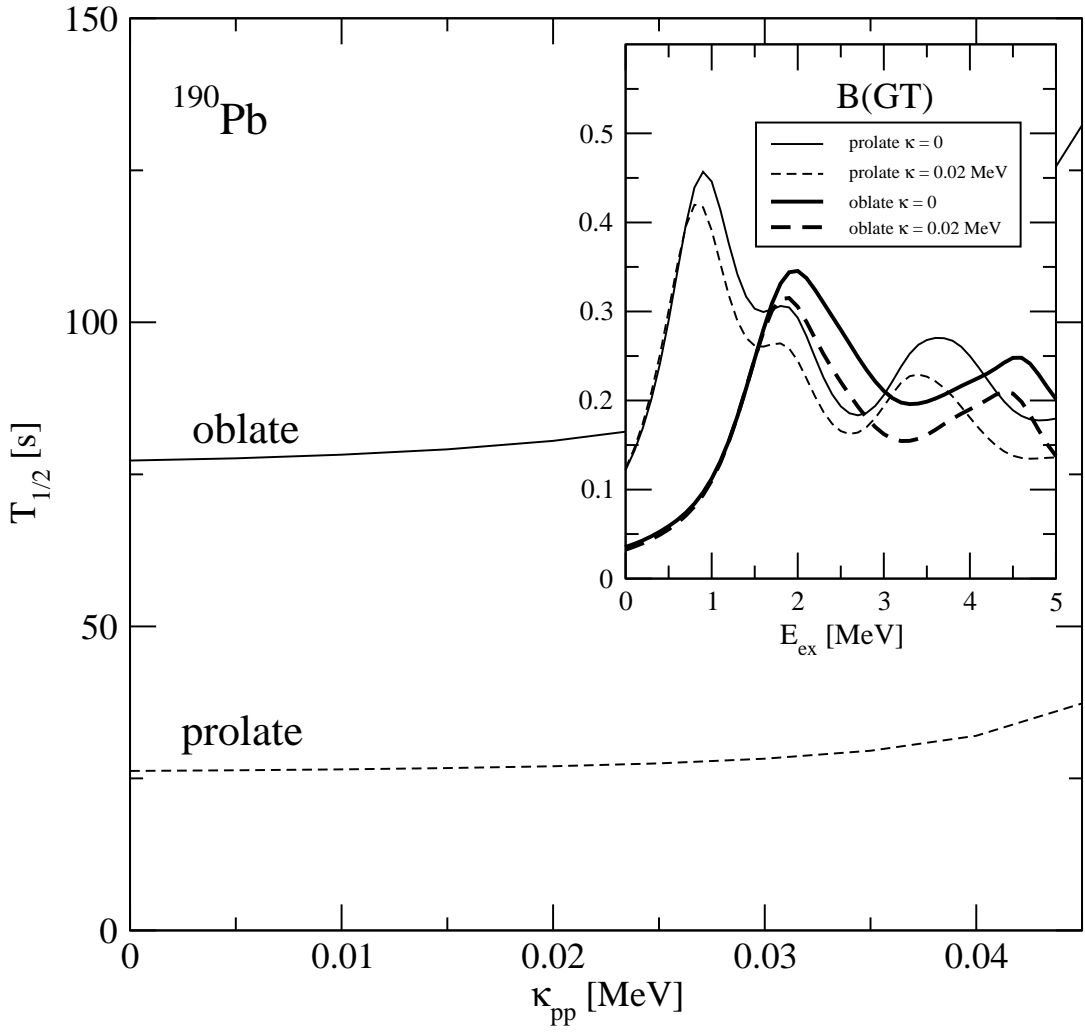}
\vskip 1cm
\caption{Half-lives  for the oblate and prolate shapes of $^{190}$Pb as a 
function of the coupling strength of the particle-particle residual 
interaction $\kappa_{pp}$. In the inset we can see the $B(GT)$ strength as
a function of the excitation energy for various shapes and $\kappa_{pp}$
parameters.}
\end{figure}

\newpage

\begin{figure}[t]
\epsfig{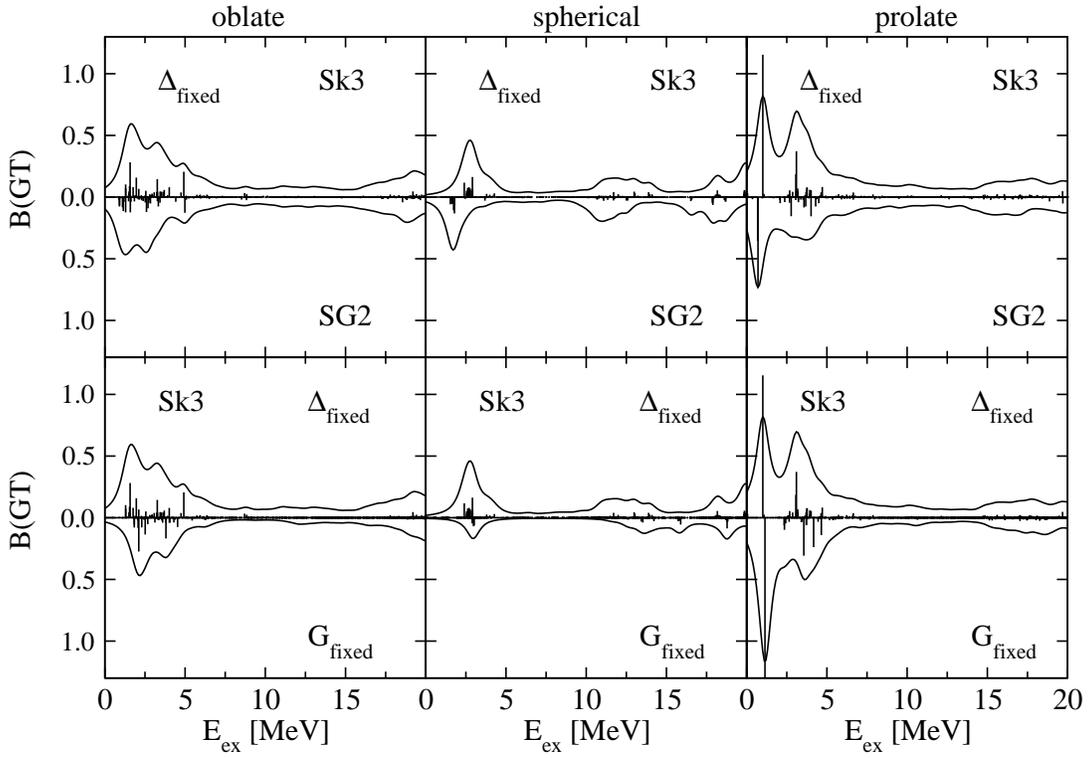}
\vskip 1cm
\caption{Gamow-Teller strength distributions in $^{184}$Pb as a function
of the excitation energy in the daughter nucleus. From left to right panels 
we show the results for oblate, spherical, and prolate shapes. In the upper
panels we show the results from Sk3 (upward) and SG2 (downward) 
with fixed gap parameters, while in the lower panels we show the results 
from fixed pairing gaps (upward) and constant pairing strengths (downward) 
for Sk3.}
\end{figure}

\newpage

\begin{figure}[t]
\epsfig{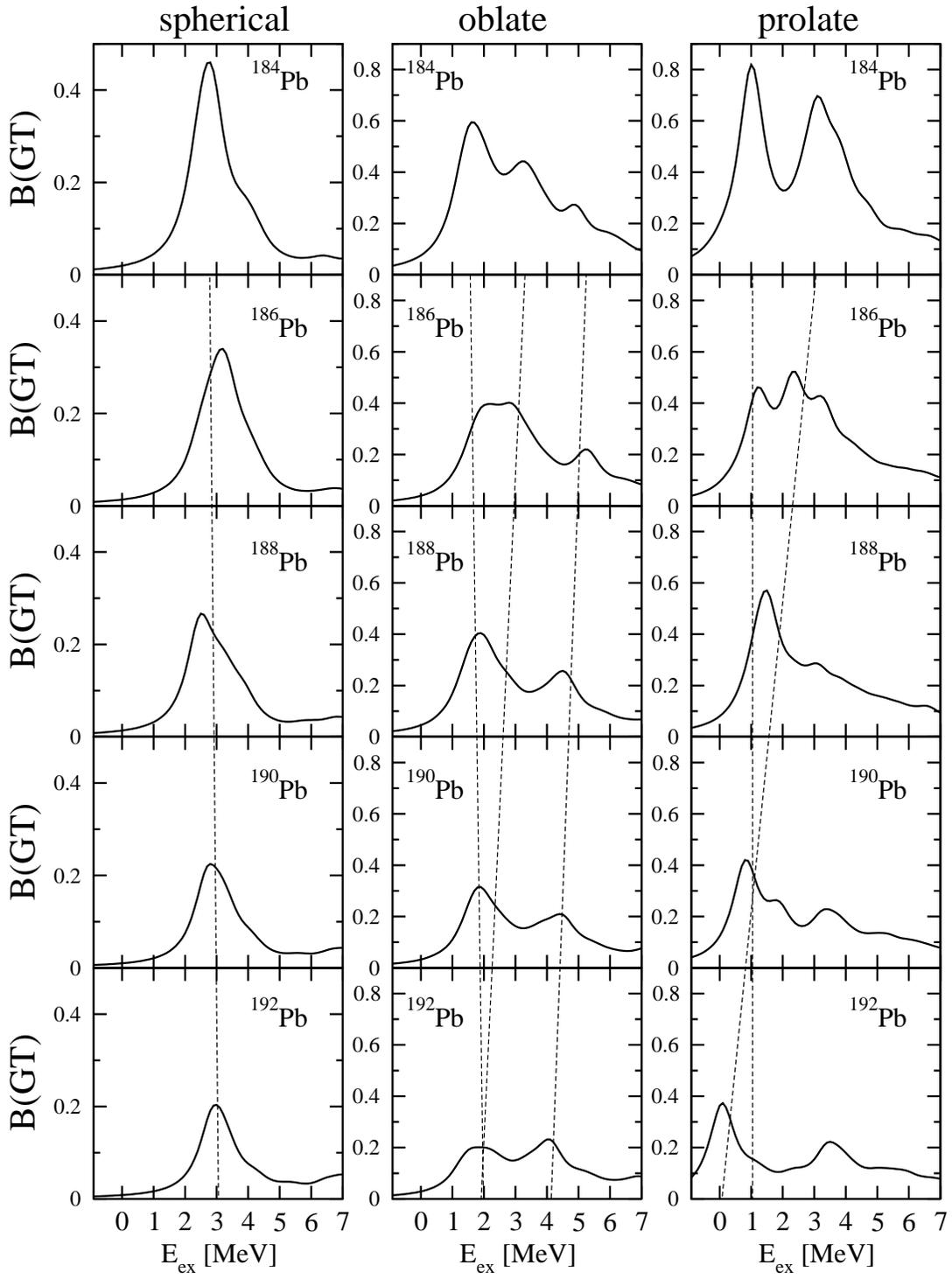}
\vskip 1cm
\caption{Gamow-Teller strength distributions in the
$^{184,186,188,190,192}$Pb isotopes for spherical (left),
oblate (middle) and prolate (right) shapes. Results are obtained
from Sk3 force with fixed gap parameters.}
\end{figure}

\newpage

\begin{figure}[t]
\epsfig{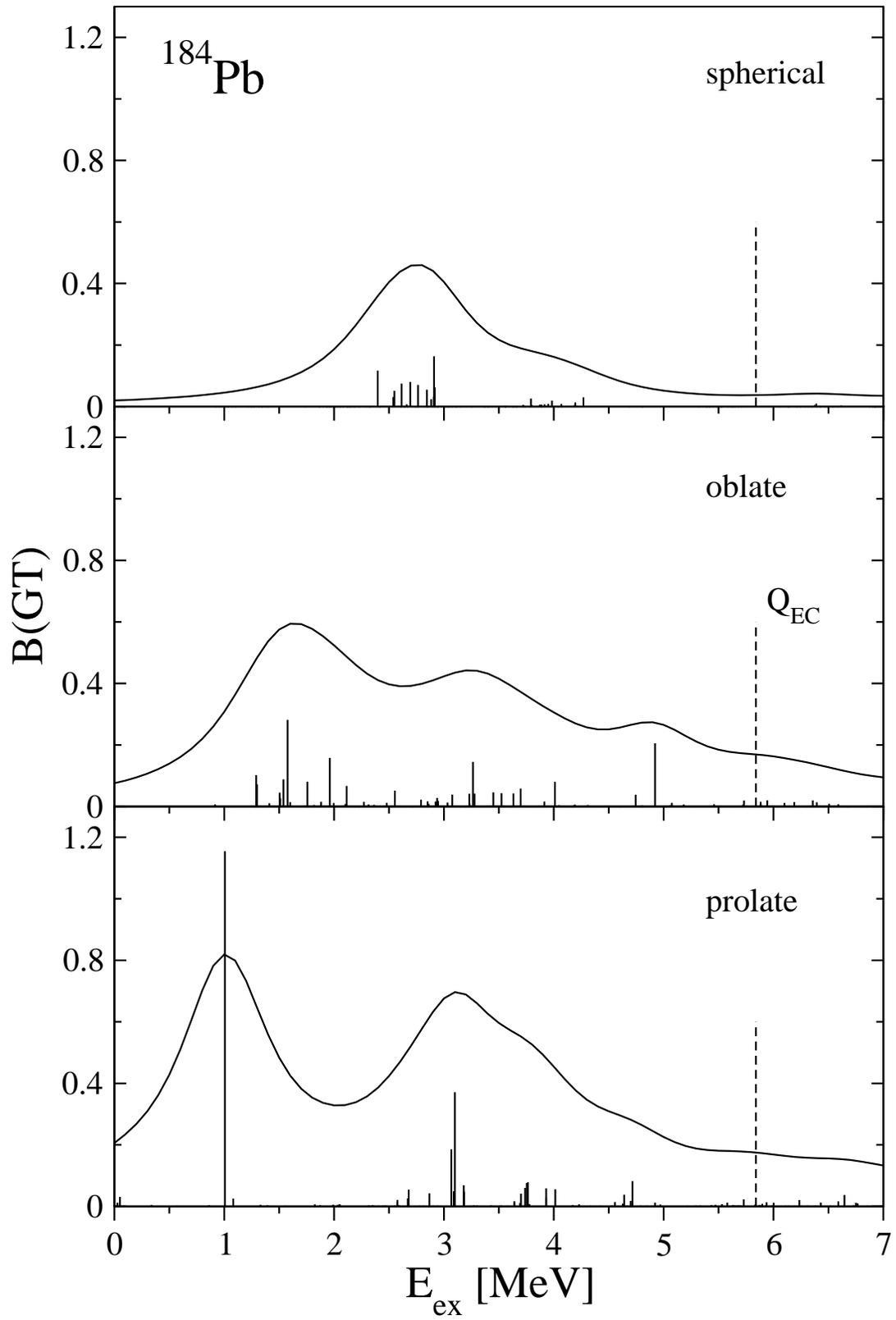}
\vskip 1cm
\caption{Gamow-Teller strength distributions (discrete and folded) in 
$^{184}$Pb. The experimental $Q_{EC}$ energy is shown with a dashed 
vertical line.}
\end{figure}

\newpage

\begin{figure}[t]
\epsfig{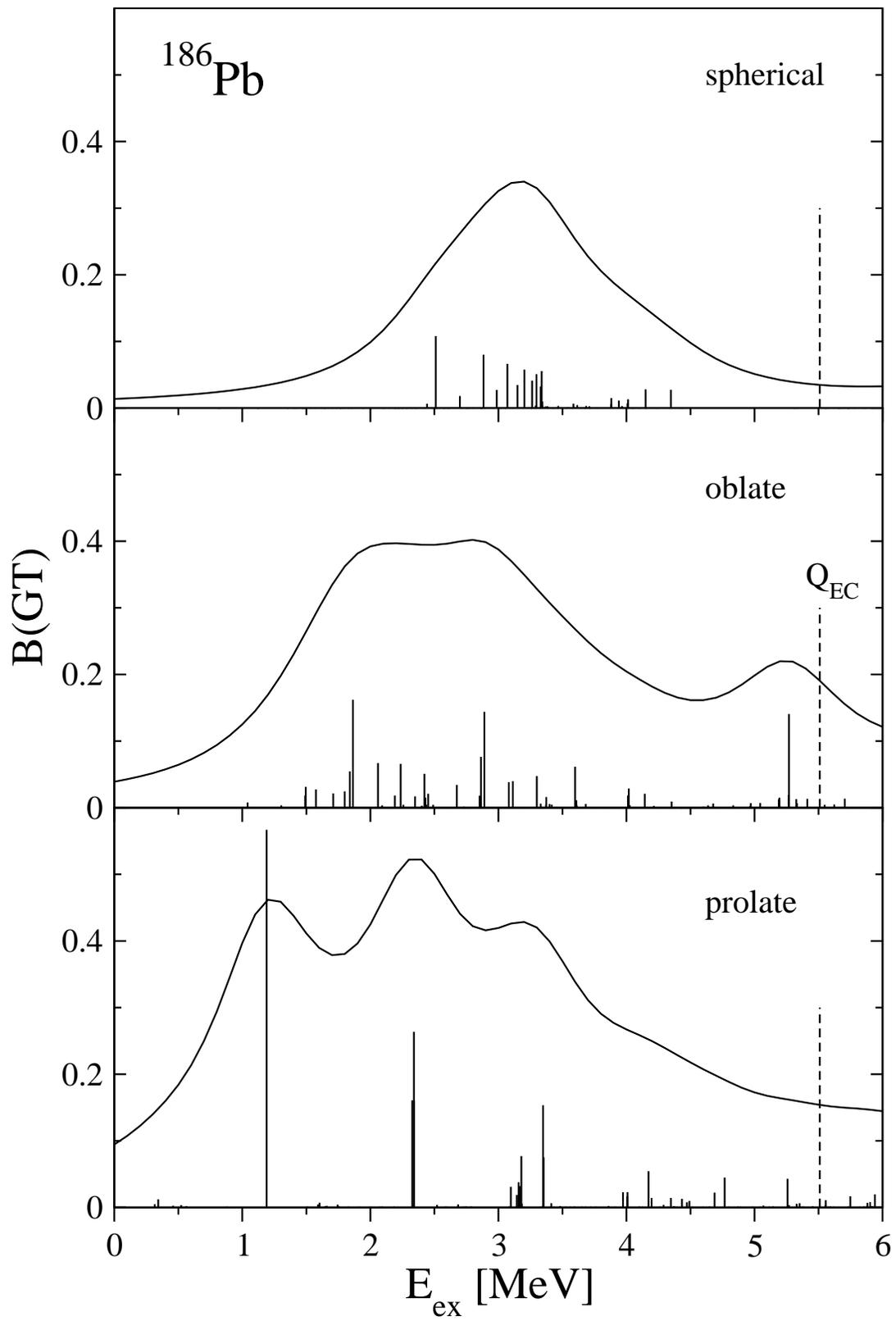}
\vskip 1cm
\caption{Same as in Fig. 6 for $^{186}$Pb.}
\end{figure}

\newpage

\begin{figure}[t]
\epsfig{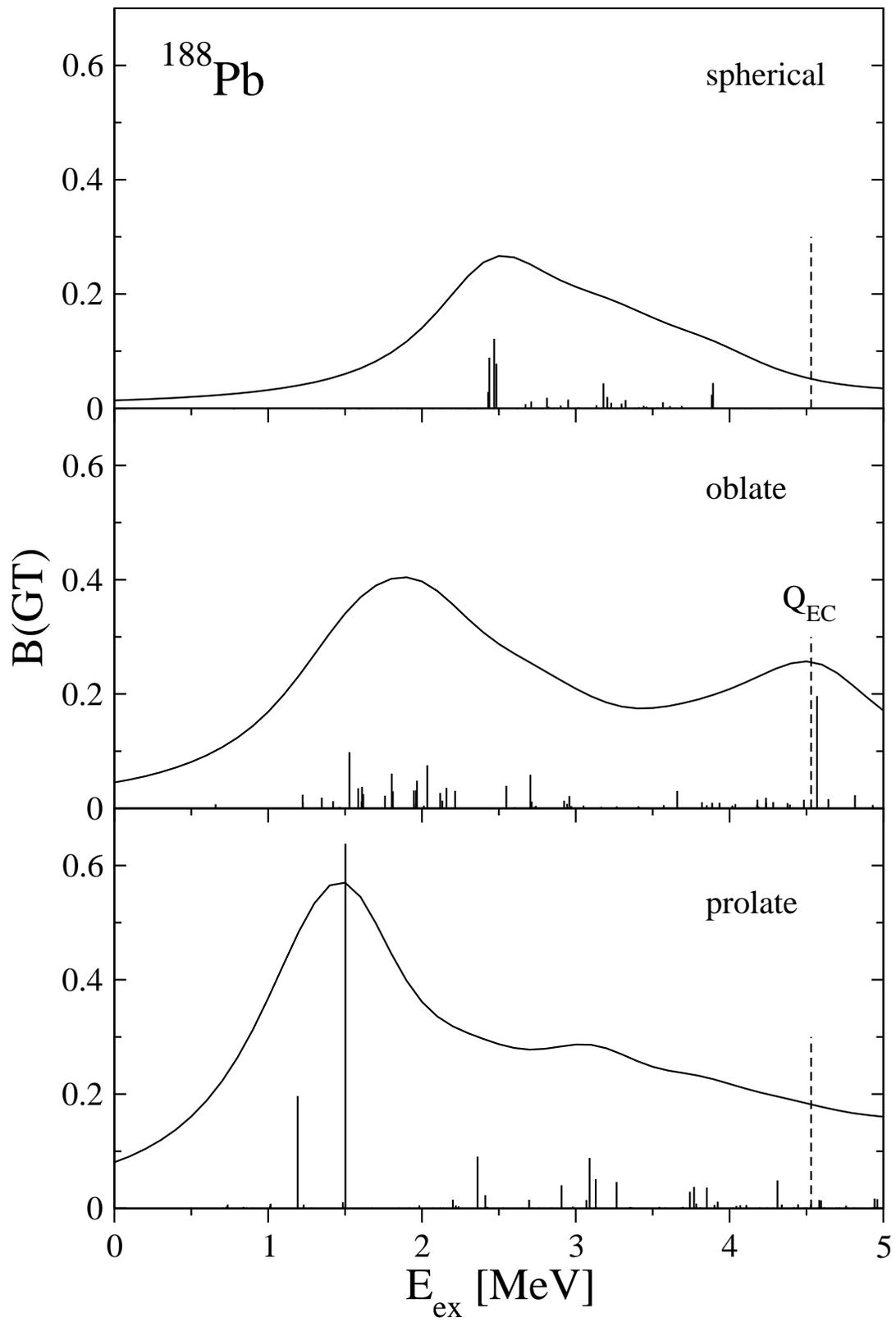}
\vskip 1cm
\caption{Same as in Fig. 6 for $^{188}$Pb.}
\end{figure}

\newpage

\begin{figure}[t]
\epsfig{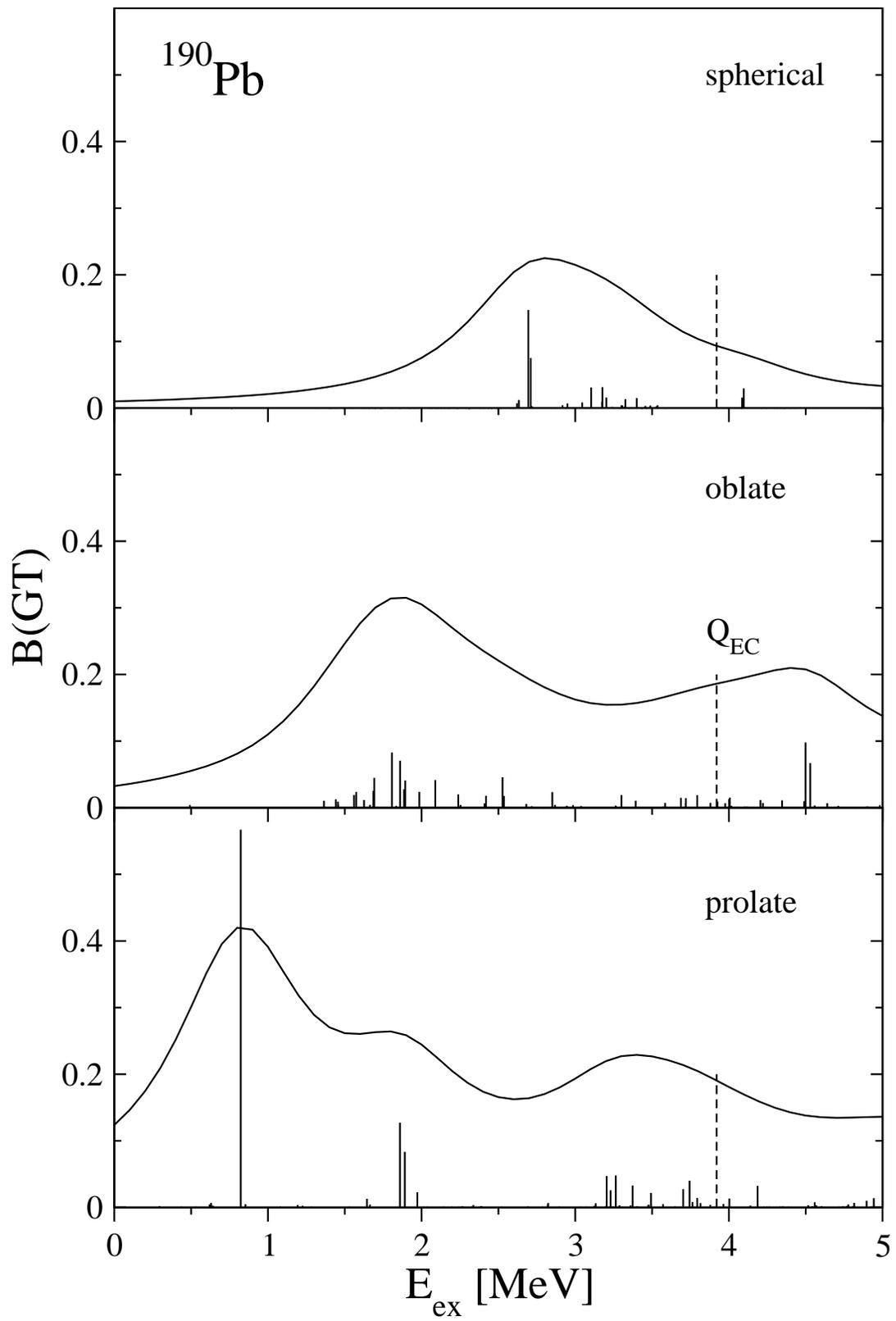}
\vskip 1cm
\caption{Same as in Fig. 6 for $^{190}$Pb.}
\end{figure}

\newpage

\begin{figure}[t]
\epsfig{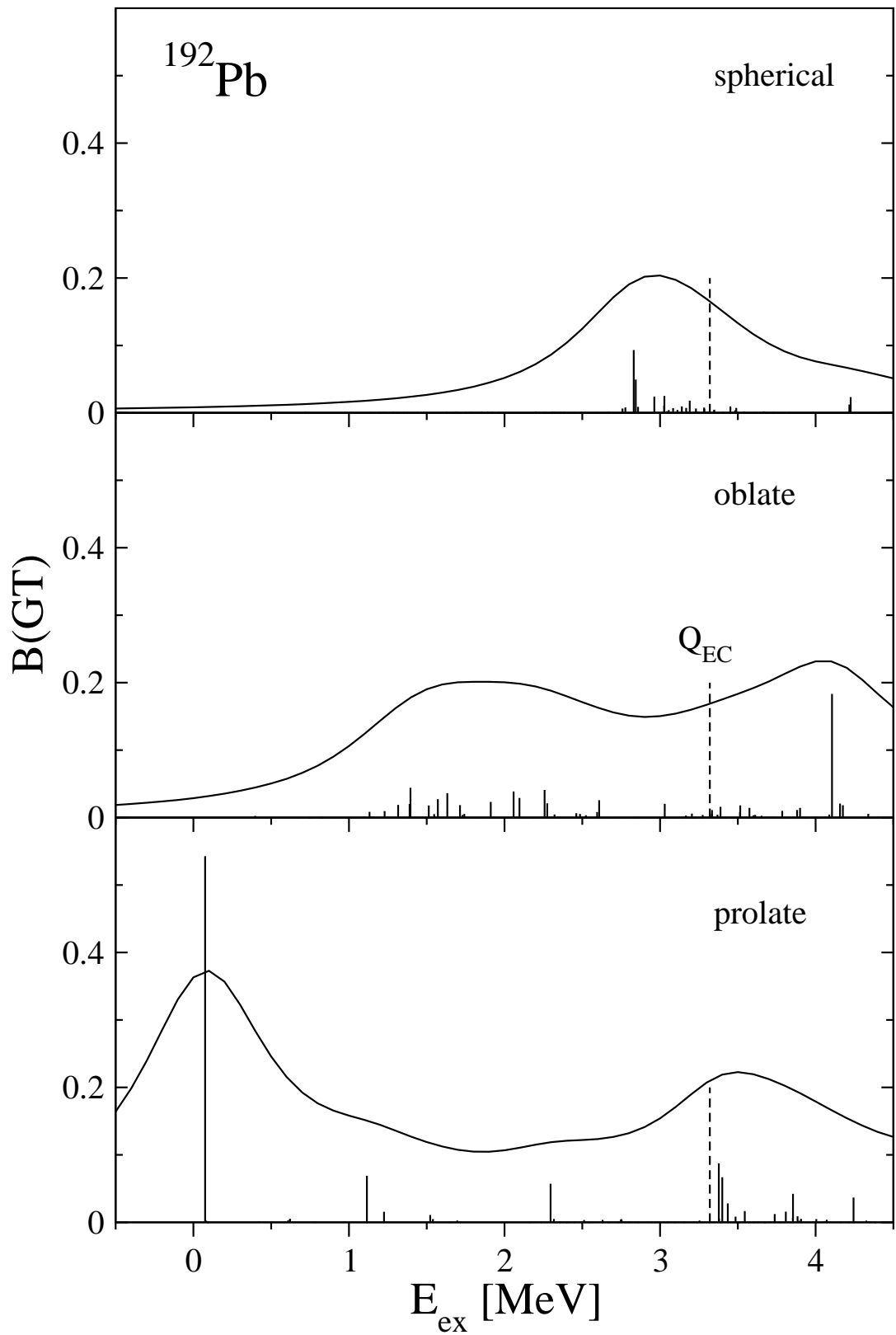}
\vskip 1cm
\caption{Same as in Fig. 6 for $^{192}$Pb.}
\end{figure}

\newpage

\begin{figure}[t]
\epsfig{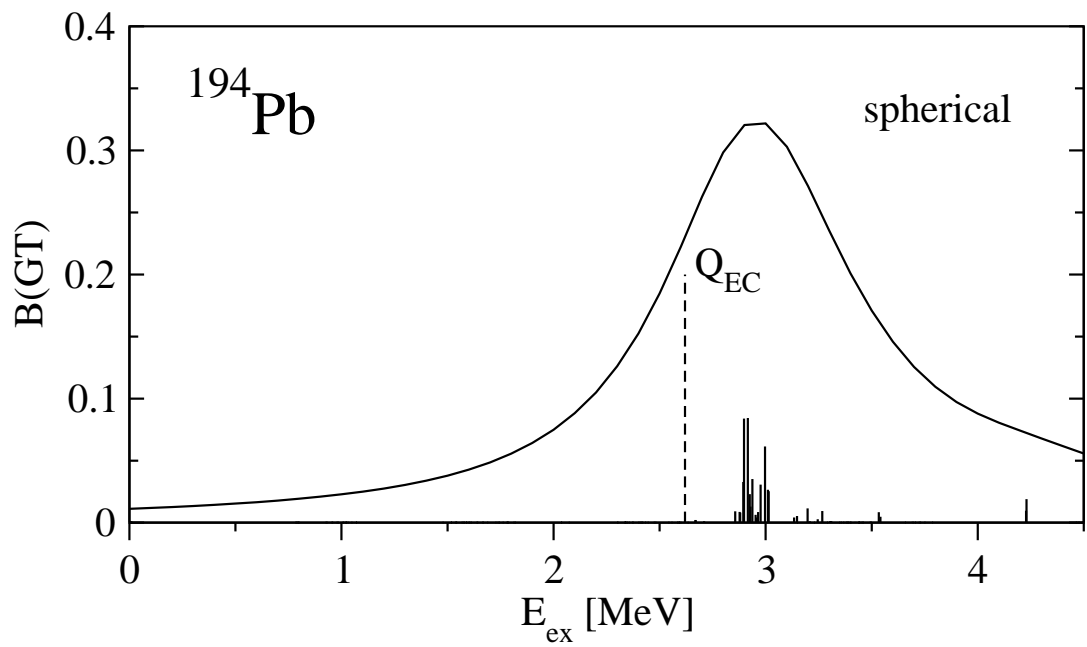}
\vskip 1cm
\caption{Same as in Fig. 6 for $^{194}$Pb.}
\end{figure}

\end{center}

\end{document}